\documentclass{article}

\usepackage{arxiv}

\usepackage[utf8]{inputenc} % allow utf-8 input
\usepackage[T1]{fontenc}    % use 8-bit T1 fonts
\usepackage{hyperref}       % hyperlinks
\usepackage{url}            % simple URL typesetting
\usepackage{booktabs}       % professional-quality tables
\usepackage{amsfonts}       % blackboard math symbols
\usepackage{nicefrac}       % compact symbols for 1/2, etc.
\usepackage{microtype}      % microtypography
\usepackage{lipsum}
\usepackage{graphicx}
\graphicspath{ {./Figures/} }

\usepackage{graphicx}
\usepackage{amsmath}
\usepackage{amssymb}
\usepackage{booktabs}
\usepackage{longtable}
\usepackage{acronym}

\newcommand{\gan}{CGAN-ECT }
\usepackage{textcomp}
\usepackage{wrapfig}
\usepackage[caption=false]{subfig}
\usepackage{makecell, multirow}
\usepackage[export]{adjustbox}
\usepackage{kantlipsum}
\usepackage{booktabs}
\usepackage{color,soul}
\usepackage{float}

\usepackage{algorithm}
\usepackage{algpseudocode}

\title{CGAN-ECT: Tomography Image Reconstruction from Electrical Capacitance Measurements Using CGANs}

\author{Wael Deabes\\
Department of Computational, Engineering, Mathematical Sciences (CEMS)\\
Texas A\&M University-San Antonio\\
One University Way, HALL 334\\
San Antonio, TX  78224\\
{\tt\small wdeabes@tamusa.edu}
\And
Alaa E. Abdel-Hakim\\
Computer Science Dep. in Jamoum,\\ Umm Al-Qura University\\
Electrical Engineering Department\\
Assiut University\\
Assiut, Egypt, 71516\\
{\tt\small adali@uqu.edu.sa;}\\
{\tt\small alaa.aly@eng.au.edu.eg}
}

\begin{document}
\maketitle
\begin{abstract}
% \todo{
Due to the rapid growth of Electrical Capacitance Tomography (ECT) applications in several industrial fields, there is a crucial need for developing high quality, yet fast, methodologies of image reconstruction from raw capacitance measurements. Deep learning, as an effective non-linear mapping tool for complicated functions, has been going viral in many fields including electrical tomography. In this paper, we propose a Conditional Generative Adversarial Network (CGAN) model for reconstructing ECT images from capacitance measurements. The initial image of the CGAN model is constructed from the capacitance measurement. To our knowledge, this is the first time to represent the capacitance measurements in an image form. We have created a new massive ECT dataset of 320K synthetic image-measurements pairs for training, and testing the proposed model. The feasibility and generalization ability of the proposed \gan model are evaluated using testing dataset, contaminated data and flow patterns that are not exposed to the model during the training phase. The evaluation results prove that the proposed \gan model can efficiently create more accurate ECT images than traditional and other deep learning-based image reconstruction algorithms. \gan achieved an average image correlation coefficient of more than $99.3\%$ and an average relative image error about $0.07$.
% } % \todo
\end{abstract}

\section{Introduction}
\label{sec:intro}

\begin{figure}[!th]
    \centering
    \includegraphics[width=.6\textwidth]{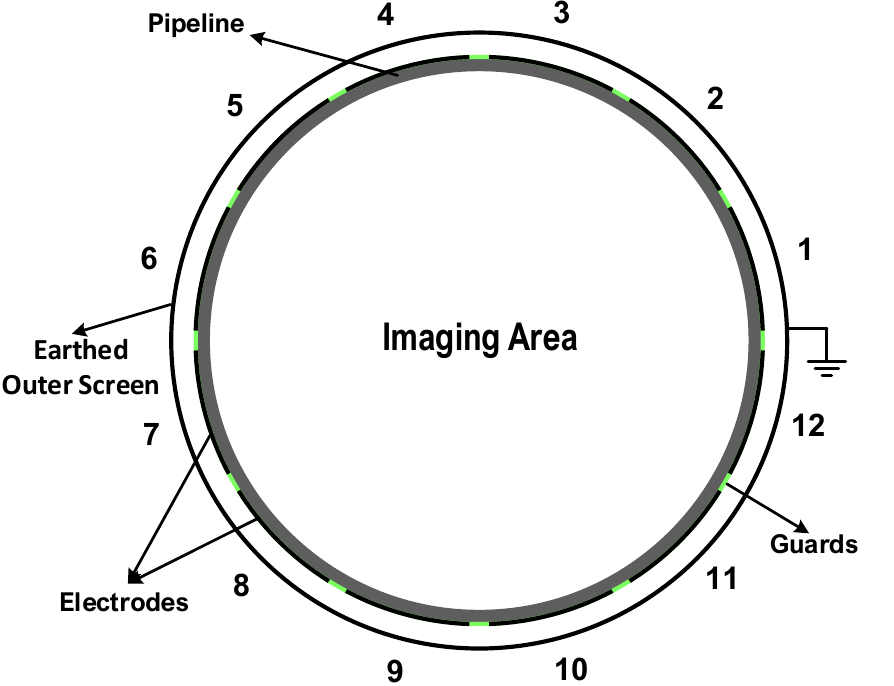}
    \caption{ECT sensor with 12 electrodes}
    \label{fig:Sensor}
\end{figure}

Tomography plays a crucial role in industrial fields, e.g. food industry, industrial tomography, biomedical~\cite{korjenevsky2004electric}, chemical, and pharmaceutical processes~\cite{wang2015industrial, Wang_2011}, and non-destructive assessment of invisible objects in dams and flood embankments~\cite{rymarczyk2018use,rymarczyk2018non}. The inner distribution of materials inside pipes is essential in such application domains. Electrical Tomography (ET) is an imaging technology that can provide such distributions in a form of cross-sectional images using non-invasive outer electrical measurements~\cite{AbdRashid2016}. Capacitance is a proper electrical property that fits this purpose. The outer capacitance measurement can capture the variations in  permittivity distributions of the inner materials. This is achieved by Electrical Capacitance Tomography (ECT). The ECT is an imaging modality that enjoys several advantages over other tomography modalities. Beside it is non-invasive and non-destructive, it is fast, inexpensive, and radiation hazards-free. Moreover, various electrical properties of diverse flow materials can be represented by ECT.

% The operation principle of ECT depends on capturing the variations in the capacitance, which is produced by permittivity changes of the inner materials.
% This capacitance changes induce variations in the outer capacitance measurements reflecting the inner distribution of materials~\cite{AbdRashid2016}.
% ECT generates distribution images based on the outer capacitance measurements~\cite{AbdRashid2016}. Accordingly,
An ECT system contains in its core an array of capacitance sensors, along with a data acquisition hardware, and a post-processing unit, which is typically an imaging computer. Multiple exciting electrodes are evenly mounted around the imaged vessel in order to build up an ECT sensor~\cite{Seong2015HardwareMeasurements}, as shown in Fig. \ref{fig:Sensor}. With the aid of these sensors, the ECT system can determine the changes in the capacitance measurements due to variations of dialectic material distribution inside the imaging area. The collected measurements are used to build cross-sectional images representing the permittivity distribution.
% ~\cite{Deabes2020}.
This process is called the inverse problem and is accomplished using image reconstruction algorithms~\cite{Cui2016,Deabes16}.

The ECT image reconstruction problem is being paid special interest in literature. This is due to its sophistication resulted from the non-linear relationship between the permittivity distribution and the measurements. Additionally, from a practical perspective, the choice of the optimal imaging method for a specific application is not straightforward.

The ECT image reconstruction process can be performed using non-iterative algorithms, e.g. Linear Back Projection (LBP)~\cite{gamio2005electrical}; or iterative algorithms, e.g. Tikhonov regularization~\cite{Vauhkonen1998TikhonovTomography}, Newton Raphson~\cite{chen2009novel}, and Iterative Landweber Method (ILM)~\cite{li2008image}. Although the non-iterative methods are obviously fast and simple, their reconstructed images usually suffer from deformations. On the other side, iterative methods generate images with higher quality. However, they are computationally expensive. Therefore, they are more suitable for offline applications.

Machine Learning (ML), particularly Deep Learning (DL), methods showed up as the tools that can compromise the tradeoff between reconstructing high-quality images and computation efficiency. The ability of DL to map complex non-linear functions enables them to be endorsed by many domains~\cite{Zhu2021}. Several attempts have been made to exploit ML in ECT. A non-linear image reconstruction approach was evolved using a conjoined Feed Forward Neural Network (FFNN) and analog Hopfield Network~\cite{Marashdeh2006}. Deabes et al.~\cite{Deabes2021}, implemented a highly reliable (ECT-LSTM-RNN) model for industrial applications to predict the metal distribution during the Lost Foam Casting (LFC) process. Also, they developed a LSTM-Image Reconstruction (LSTM-IR) algorithm ~\cite{deabes2021image} to generate high quality ECT images. Zhu et a.~\cite{Zhu2020} presented a forward, an inverse, permittivity prediction network in a deep neural network to reconstruct the object shape. For the improvement purposes of the reconstructed images of multiple different permittivity values in the sensing field, Zhu et al. added multi-level fusion layer to DL conventional methods~\cite{Zhu2021}. In~\cite{deabes2022adversarial}, an adversarial resolution enhancement (ARE-ECT) model was presented to reconstruct high-resolution images of inner distributions based on low-quality initial images. A Residual Autoencoder called (ECT-ResAE) was implemented in~\cite{deabes2022residual} to generate ECT images from modulated capacitance measurement images. 

Despite the success achieved by such methods, they depend on small training datasets. This leads to generalization problems, which appears with never-seen capacitance vectors. So, generating large-scale ECT datasets for training, validation, and testing purposes along with and implementing new ML image reconstruction frameworks are essential for better generalization capabilities in extensive applications. In this paper, a DL framework called \gan model for creating tomography images from capacitance measurements is developed. The model uses Conditional Generative Adversarial Networks (GAN)~\cite{mirza2014conditional} with a UNet generator~\cite{ronneberger2015u}. 
The GANs have yielded splendid results in many image generating tasks, such as photographs of human faces, image-to-image Translation, photograph editing, and others \cite{radford2015unsupervised}. The inputs of the proposed \gan model are the capacitance measurements after a modulation process converting it to a square matrix. The output is a reconstructed tomography image identifying all the objects inside the imaging area. The trained \gan model is obviously fast since its forward operation mode. A large-scale dataset with $320k$ pairs has been created to train and test the proposed model. The proposed method yields more accurate images than other explicit approaches with a lower computational cost.

\begin{figure*}[!thp]
    \centering
    \includegraphics[width=.9\linewidth]{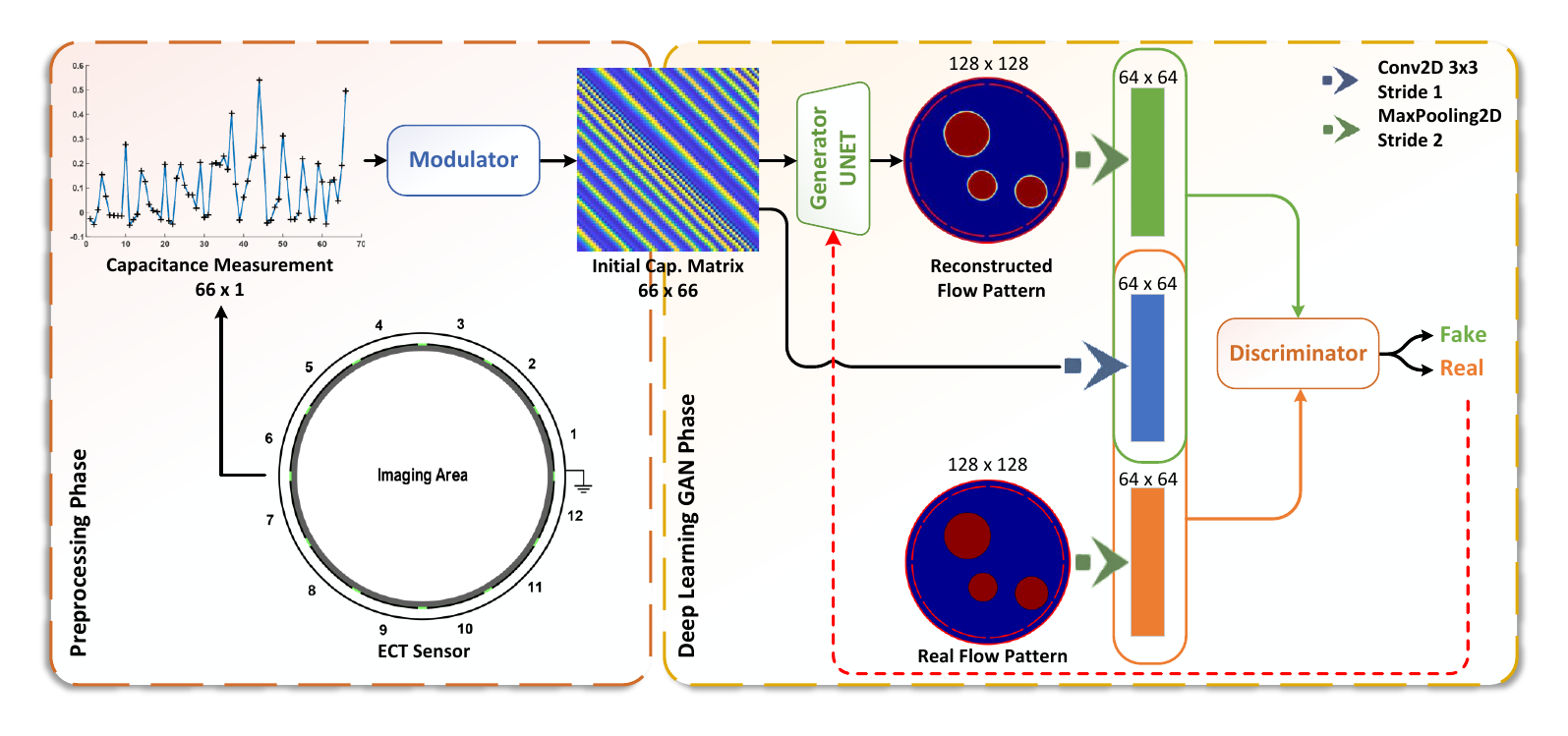}
    \caption{Architecture of \gan model}
    \label{fig:GANArch}
\end{figure*}

\section{\gan Model Description}
In ECT, the relationship between capacitance measurement readings and cross-sectional images is formulated as a two-way mapping problem. The forward problem is relatively the easier one. It is concerned by estimating the capacitance readings given cross-sectional data, as shown in Eq.~\ref{eq.3}.

\begin{equation}
    C_{M \times 1} = S_{M \times N}(\varepsilon_0) \cdot G_{N \times 1}
\label{eq.3}
\end{equation} 
where $S$ is the sensitivity matrix, which is the Jacobian of the capacitance with respect to pixels evaluated at $\varepsilon_0$ (the first permittivity distribution), $N$ is the number of image's pixels, $M$ is the number of sensor-pair readings. $G$ is the permittivity distribution, and $C$ is the calculated capacitance.

The inverse problem is simply estimating $G$ given $C$. Obviously, the inverse problem is an ill-posed one. Therefore, the generated cross-sectional image is very sensitive to measurement errors~\cite{Isaksen1996a}. The existing algorithms for the reverse problem usually suffer from quality or efficiency drawbacks. We propose the \gan model to overcome these ECT related problems. The following subsections explain the main components of the proposed \gan model, as shown in Fig. \ref{fig:GANArch}.

\subsection{Modulator}
The ECT capacitance sensor produces a $1\times 66$ raw vector data, i.e. $M=66$. If the sensor set is rotated CCW, another $1\times 66$ vector is generated. This generated vector is rotate-right version of the first one. When this single step rotation is repeated, a matrix of $66\times 66$ is generated. Such kind of sensor virtual rotation achieves the  rotation-invariance property for the inner material distribution. In other words, the generated matrix is rotation invariant. This property is extremely crucial, specially for the case of asymmetric flow patterns. Rotation invariance achieves compactness of the input data with diverse poses of inner objects. Particularly, for an asymmetric flow pattern, a single matrix provides comprehensive readings regardless the relative placement of the sensors with respect to the inner pattern spatial orientation. The main objective of the modulator is to implement such readings augmentation. Specifically, it is responsible of generating the rotation-invariant initial capacitance matrix from a single vector reading. Algorithm~\ref{alg:modulator} shows the operation of the modulator. 
    
\begin{algorithm}
\caption{The modulator operation}\label{alg:modulator}
    \begin{algorithmic}
    
        \Procedure{modulator}{$C_{1\times M},C_{M\times M}$}
            \State $v\gets C_{1\times M}$
            \State $C_{M\times M}\gets zeros(M,M)$
            \State \For{$i=1:M$} 
                    \State $C_{M\times M}[i,:]\gets [v]$
                    \State $v\gets rotate\_right(v,1)$
            \EndFor
                    
            \State \textbf{return} $C_{M\times M}$
        \EndProcedure
          
    \end{algorithmic}
   
\end{algorithm}

% \vspace{-12pt}  
\subsection{Generator}
The spatial information of the inner material distribution is crucial in constructing the image of the flow pattern. Although traditional autoencoders are good in reconstructing such patterns, they cannot recover the spatial information contents with sufficient accuracy. UNet has been proven to cover this gap~\cite{ronneberger2015u}. Therefore, we use UNet in the generator to construct the final flow pattern. We use $4$ blocks in the encoder side, and $4$ blocks in the decoder side. The size of the latent vector is $8$. The initial modulated capacitance matrix of the input layer is concatenated with the generated image from final layer. Similarly, each input of the hidden layers in the decoder side is concatenated with the output of the corresponding layer from the encoder side.
    
\subsection{CGAN Model}
The core of the deep learning phase is a CGAN network~\cite{mirza2014conditional}. The constructed flow patterns that are generated by the UNet generator, $G(C_{M\times M})$, work as the fake samples. The real samples, $FP_r$, are obtained from the synthetic image set, $\mathcal{FP}$, which are generated by the proposed synthetic generator described in section~\ref{sec:Data}. The generated image from the generator, and the real image are passed through a maxpooling layer with $64$ filters, while the initial modulated capacitance matrix is convoluted by a Conv2D layer with a $3 \times 3$ mask and a stride of $1$ to generate a downsized image of size equals $64 \times 64$. Both real and fake samples are conditioned by that produced  $64\times 64$ image, as shown in Eq.~\ref{eq:cgan}.

\vspace{-10pt}
\begin{equation}
\label{eq:cgan}
\begin{aligned}
    G^* = &\arg\min _G\max_ D(\underset{FP_r\in\mathcal{FP}}{\mathbb{E}}[\log (D(FP_r|C_{M\times M}))]+\\
    &\underset{C_{M\times M}\in\mathcal{CM}}{\mathbb{E}}[\log(1-D(G(C_{M\times M})|C_{M\times M}))])
\end{aligned}
\end{equation}
\vspace{-2pt}
\noindent where $\mathcal{CM}$ is the set of capacitance measurement matrices.

\section{Experimentation}
\subsection{Experimental Setup and ECT Dataset}
\label{sec:Data}
\begin{figure*}[!thp]
    \centering
    \includegraphics[width=.7\linewidth]{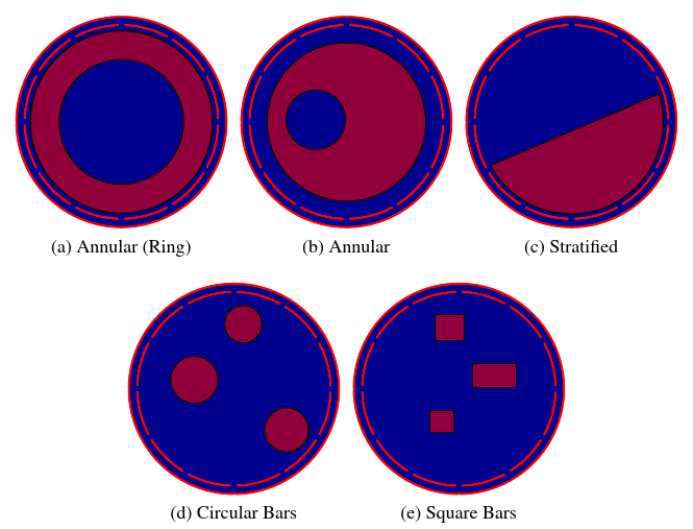}
    \caption{Samples of different flow patterns}
    \label{fig:Patterns}
\end{figure*}

We have implemented a Matlab GUI software package to realize different configurations of ECT sensors. By solving the ECT forward problem, various flow patterns can be simulated, and their corresponding capacitance measurements are generated. Accordingly, an extensive ECT benchmark dataset has been developed to train and test the proposed algorithm. The dataset consists of $320k$ samples, each one is a pair of an actual permittivity distribution vector, and its corresponding capacitance measurements modulated matrix. The size of the actual flow pattern is $128\times 128$, while the modulated matrix is $66 \times 66$. The ECT sensor consists of $12$ electrodes as shown in Fig. \ref{fig:Sensor}. The sensor pipe is made from PVC material with a relative permittivity of $2$. The diameter and the thickness of the pipe are $93mm$ and $2.5mm$, respectively. The electrodes are separated by gaps of $4^o$, and the span angle of each electrode is $26^o$. The dataset contains $9$ different flow patterns: $10k$ ring patterns, annular with $20k$ patterns, $10k$ stratified patterns, $1-3$ circular bars with $140k$ patterns, and $140k$ patterns of $1-3$ square bars, as shown in Fig. \ref{fig:Patterns}. The low permittivity phase is air with value $1$, and the high phase is plastic particles with permittivity of $4$. the physical specifications of every pattern are assigned stochastically. For instance, the ring's width for annular flow is assigned to a random variable between $10\%$ and $95\%$ of its radius. The height of the stratified flow is randomly set between $5\%$ and $95\%$ of the diameter. The number of bars varies from $1$ to $3$. Additionally, for testing, the trained \gan  model, a real ECT system is used to measure real capacitance data. 

\subsection{Evaluation Metrics}
In the ECT inverse problem, usually the image quality, and reconstruction algorithm’s performance are evaluated using two common metrics: the relative Image Error (IE) and Correlation Coefficient (CC) between the actual and the reconstructed flow patterns~\cite{Cui2016}. The relative IE is defined as shown in Eq.~\ref{norm1}. 
\begin{equation}
IE = \frac{||G-G^*||_2}{||G||_2} 
\label{norm1}
\end{equation}
where $G^*$ represents the predicted permittivity distribution from the generator network of \gan image reconstruction model, and $G$ represents the original material distribution.

The CC, which is shown in Eq.~\ref{Eq.CC}, measures how similar the reconstructed image is to the actual flow distribution.   
\begin{equation}
CC = \frac{\sum_{i=1}^N (G_i - \bar{G})(G^*_i - \bar{G^*})}{\sqrt{\sum_{i=1}^N (G_i- \bar{G})^2  \sum_{i=1}^N(G^*_i - \bar{G^*})^2}} 
\label{Eq.CC}
\end{equation}
where $G$ is the actual permittivity distribution of the test object, $G^*$ is the reconstructed permittivity distribution, and $\bar{G}$ and $\bar{G^*}$ are the mean values of $G$ and $G^*$, respectively. $N = 16384$ is the number of pixels in the imaging area. 

The training and testing process of the \gan model is implemented in Python on TensorFlow machine learning platform \cite{tensorflow2015}, and Keras deep learning API \cite{keras2015}. When the proposed model is tested, the input is the modulated capacitance matrix, while the reconstructed permittivity flow pattern is the output. Since the testing set consists of $96k$ samples, therefore the performance is measured based on the mean values of these two metrics. Smaller relative IE and larger CC correspond to better performance. 
%%%%%%%%%%%%%%%%%%%%%%%%%%
\subsection{Evaluation Results}
The overall network's performance of the proposed model is verified based on the reconstruction results of the testing dataset. The more comprehensive the data simulation, the stronger the generalization performance of the model after training. Therefore, the generalization ability of the proposed model is tested by using testing dataset, generated phantoms, which are not included in the training dataset, and real experimental data.

\subsubsection{Qualitative Results on Simulation Test Dataset}
\begin{figure}%[!htp]
    \centering
    \includegraphics[width=.7\linewidth]{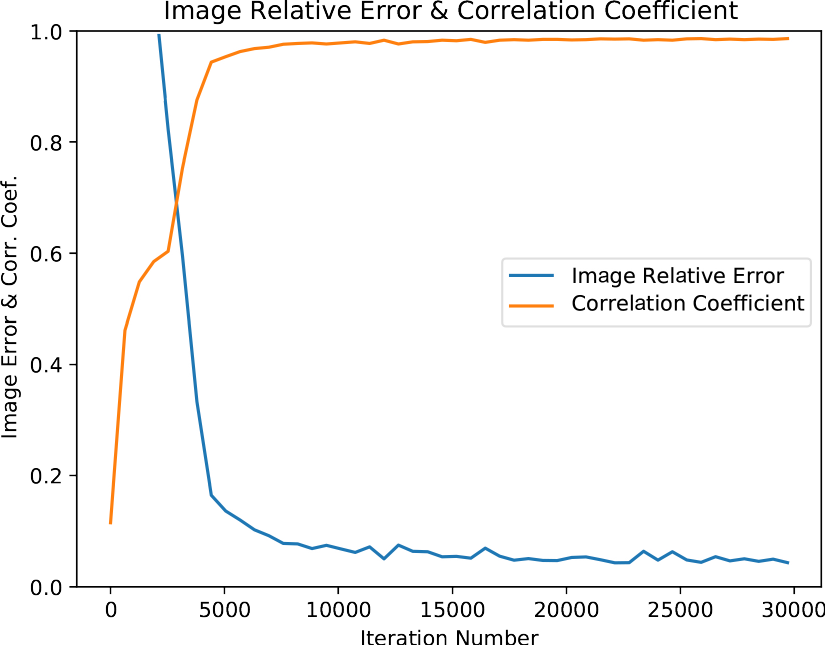}
    \caption{Training and validation loss curves}
    \label{fig:TVLC}
\end{figure}

Practically, a simulation testing dataset, which had not been exhibited to the network during the training process, is used to verify the reconstruction ability of the proposed \gan model. Typically, the ECT dataset containing $320k$ pairs is divided into $70\%~(224k~pairs)$ for training, and $30\% ~(96k~pairs)$ for testing. The training and testing datasets are quite different, since the dataset for each flow pattern was randomly generated. During the training phase, the testing dataset was applied to the \gan model and their values were calculated in each epoch. Fig.~\ref{fig:TVLC} shows that the IE curve declines over $30k$ epochs, while the CC curve increases. Both curves approaches their saturation levels, which are $99\%$ and $0.07$ for CC and IE, respectively, after about $10k$ epochs.

\setlength{\tabcolsep}{4pt}
\renewcommand{\arraystretch}{1.2}
\begin{table*}[!htp]\centering
\caption{Minimum and maximum of relative IE and CC of testing results}
\label{Tab:MnMxIECC}
  \begin{tabular}{lcccccc}
    \toprule
    Flow Patterns & Min. IE & Max. IE & Average IE & Min. CC & Max. CC & Average CC\\
    \midrule
Annular            & 0.0339 & 0.1045 & 0.0593 & 0.9912 & 0.9982 & 0.9957 \\
Ring               & 0.0425 & 0.1360 & 0.0802 & 0.9835 & 0.9973 & 0.9927 \\
Stratified         & 0.0272 & 0.0761 & 0.0484 & 0.9954 & 0.9992 & 0.9973 \\
Single Cir. Bar    & 0.0457 & 0.0925 & 0.0683 & 0.9887 & 0.9983 & 0.9940 \\
Multiple Cir. Bars & 0.0448 & 0.1265 & 0.0860 & 0.9841 & 0.9980 & 0.9920 \\
Single Sq. Bar     & 0.0379 & 0.0843 & 0.0606 & 0.9890 & 0.9979 & 0.9942 \\
Multiple Sq. Bars  & 0.0372 & 0.1344 & 0.0848 & 0.9805 & 0.9982 & 0.9909 \\
    \midrule
Total Average &&&\textbf{0.0697} &&& \textbf{0.9938}\\
    % &\multicolumn{2}{c}{Total Average \ac{IE}}  & \textbf{0.1019} & \multicolumn{2}{c}{Total Average \ac{CC}}  & \textbf{0.9884} \\
    \bottomrule
  \end{tabular}
\end{table*}

\begin{figure}%[!htp]
    \centering
    \includegraphics[width=.63\linewidth]{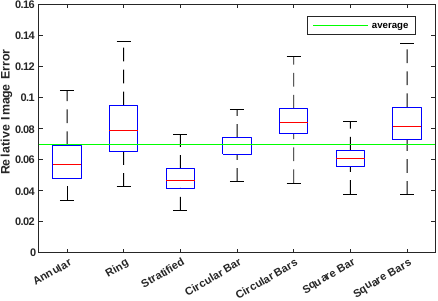}
    \caption{Box plots of Relative Image Errors (IE)}
    \label{Fig:BoxIE}
\end{figure}

\begin{figure}%[!htp]
    \centering
    \includegraphics[width=.63\linewidth]{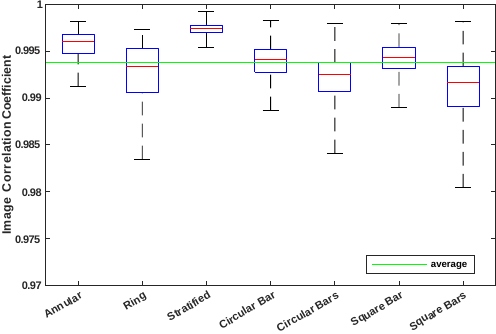}
    \caption{Box plots of Correlation Coefficients (CC)}
    \label{Fig:BoxCC}
\end{figure}

For quantitative evaluation, the minimum, the maximum, and the average of relative IE and CC are calculated for each two-phase flow patterns group of the testing dataset and shown in Table \ref{Tab:MnMxIECC}, respectively. The results prove that the proposed \gan model has reconstruction capabilities with average CC values more than $99\%$, and average IE values less than $7\%$. Box plots of the IE and CC for all flow patterns are drawn in Fig. \ref{Fig:BoxIE}, and Fig. \ref{Fig:BoxCC}, respectively. The upper and lower limits stand for the maximum and minimum of each flow pattern (exact values are given in Table \ref{Tab:MnMxIECC}), and red line is median.

% From Table \ref{Tab:MnMxIECC}, it can be found that the performance of the \gan model on multiple-bars flow are not as good as other four flow patterns. Annular flows have best results of relative IE, while as for CC, stratified and ring flows have better results. 

\begin{figure*}%[!htp]
    \centering
    \includegraphics[width=.9\linewidth]{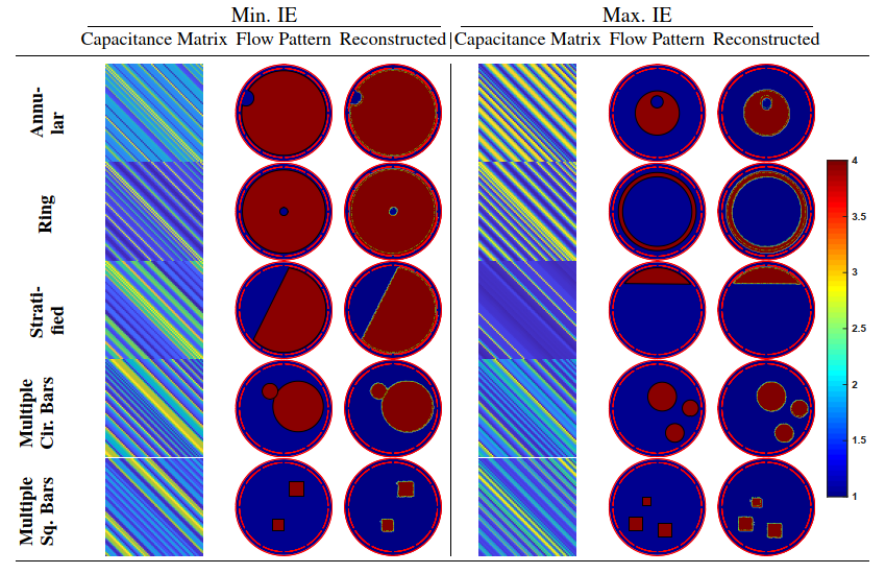}
    \caption{Examples of Min. and Max. IE image reconstruction results}
  \label{Fig:MnMxIE}
\end{figure*}

Examples of reconstructed images for the min. and max. relative image errors of each flow patterns in Table \ref{Tab:MnMxIECC} are illustrated in Fig. \ref{Fig:MnMxIE}. Visually, reconstructed images which have max. relative IEs are very close to the real flow patterns with small artifacts. The \gan model can estimate each object and identify its position in the imaging area of the ECT sensor. On the other hand, the reconstructed images with min. relative IEs have better visual effect. Also, as shown in the Fig. \ref{Fig:MnMxIE}, the images with max. relative IE are flow patterns whose low phase ratio, while those whose min. relative IE have high phase ratio, except square bars. In general, the proposed \gan model performs well on the testing dataset and has stronger image reconstruction capabilities for all typical ﬂow patterns with accurately-estimated objects permittivity values.

\begin{figure*}%[!htp]
    \centering
    \includegraphics[width=.9\linewidth]{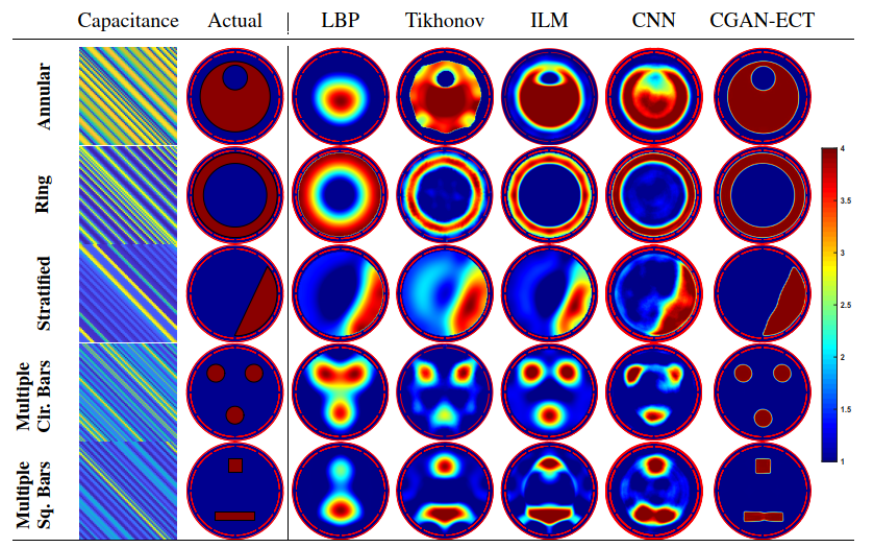}
    \caption{Reconstructed images of different state-of-the-arts image reconstruction algorithms}
  \label{Fig:Alg}
\end{figure*}

To test the performance of the proposed \gan model, a comparison with other state-of-the-arts ECT image reconstruction algorithms is carried out. The generalization ability of the \gan model is examined by applying it to a variety of flow patterns. Fig. \ref{Fig:Alg} illustrates the results where the initial modulated capacitance matrix is shown in the first column, the actual flow pattern is in the second column, the other columns have the reconstructed images from the LBP, iterative Tikhonov, ILM, CNN \cite{Zheng2019,Wang2020, Tan2018} and \gan algorithms, respectively. The iteration number of the Tikhonov algorithm is set to $200$ iterations, while the ILM is $1000$. The CNN model is trained using the initial images from the LBP algorithm. The results of the \gan model have high quality and accuracy with sharp objects’ boundaries compared with the reconstructed images from all other algorithms. Visually, it is clear from the results in Fig. \ref{Fig:Alg} that the object edges are sharp, since there is no transition zone between the reconstructed objects by the proposed model compared with the other algorithms. Also, Table \ref{Tab:Alg} contains the IE and CC results of the proposed \gan model along with all other algorithms. The values of the IE and CC  prove that \gan model outperforms the other reconstruction algorithms.

\setlength{\tabcolsep}{2.8pt}
\renewcommand{\arraystretch}{1.2}
\begin{table}[!htp]\centering
\footnotesize
\caption{Results of the IE and CC of different ECT image reconstruction algorithms}
\label{Tab:Alg}
  \begin{tabular}{lcccccccc}
    \toprule
  & Flow & LBP & Tikhonov & ILM & CNN & \gan \\
    \midrule
    \multirow{6}{4em}{\centering Relative Image Error}  & Annular     & 0.5746 & 0.3229 & 0.2693 & 0.2673 & \textbf{0.1471}\\
                                                        & Ring        & 0.3776 & 0.1216 & 0.2984 & 0.2107 & \textbf{0.0825}\\
                                                        & Stratified  & 0.2590 & 0.3793 & 0.3203 & 0.2401 & \textbf{0.1660}\\
        & \begin{tabular}{@{}c@{}}Multiple \\ Cir. Bars\end{tabular}  & 0.6083 & 0.7492 & 0.4275 & 0.4765 & \textbf{0.1340}\\
        & \begin{tabular}{@{}c@{}}Multiple \\ Sq. Bars\end{tabular}   & 0.5236 & 0.4505 & 0.4461 & 0.5933 & \textbf{0.1586}\\
    \hline
    \multirow{6}{4em}{\centering Correlation Coefficient} & Annular   & 0.4812 & 0.8421 & 0.8648 & 0.8682 & \textbf{0.9745}\\
                                                          & Ring      & 0.8110 & 0.9792 & 0.9576 & 0.9396 & \textbf{0.9907}\\
                                                        & Stratified  & 0.9126 & 0.8518 & 0.9100 & 0.9179 & \textbf{0.9719}\\
        & \begin{tabular}{@{}c@{}}Multiple \\ Cir. Bars\end{tabular}  & 0.5498 & 0.5652 & 0.7625 & 0.7325 & \textbf{0.9806}\\
        & \begin{tabular}{@{}c@{}}Multiple \\ Sq. Bars\end{tabular}   & 0.4572 & 0.7018 & 0.7238 & 0.6915 & \textbf{0.9571}\\
    \bottomrule
  \end{tabular}
\end{table}

\subsubsection{Experiments on contaminated data}
\label{ContData}
During the training and testing phases of the \gan model, contaminated noisy data was used to test its robustness. Gaussian noise with Signal to Noise Ratio (SNR) equals $30$dB was added to the capacitance vectors before the modulation process. The selected SNR is common for most of reported ECT hardware systems. The \gan model is trained and tested with noise-free, and noisy datasets. The results of the min., max., and average IE and CC for these two datasets are shown in Table \ref{Tab:NFNData}. Also, all the results of all different flow patterns groups are stated in Table \ref{Tab:NFNDataExp}. Statistically, from the results shown in the two tables, there in no big difference when applying noise-free and noisy data. However, the IE and CC values of noise-free training/ noisy testing data are slightly worse than the other models but still substantial.  
 
\setlength{\tabcolsep}{1.5pt}
\renewcommand{\arraystretch}{1.2}
\begin{table}\centering%[!htp]\centering
\caption{Results of noise-free and noisy training and testing data}
\label{Tab:NFNData}
  \begin{tabular}{lccccccc}
    \toprule
  \multicolumn{2}{c}{Training Data} & \multicolumn{2}{c}{Noise-Free} & \multicolumn{2}{c}{Noisy} \\
     \cmidrule(rl){1-2} \cmidrule(rl){3-4} \cmidrule(rl){5-6}
    \multicolumn{2}{c}{Testing Data} & Noise-Free & ~Noisy & ~Noise-Free & ~Noisy \\
    \midrule
      \multirow{3}{6em}{\centering Relative Image Error} & Min  & 0.0385 & ~0.0388 & 0.0348 & 0.0351\\
                                                         & Max  & 0.1078 & ~0.2573 & 0.1477 & 0.1829\\
                                                         & Avg. & 0.0697 & ~0.1268 & 0.0854 & 0.1001\\
    \hline
    \multirow{3}{6em}{\centering Correlation Coefficient} & Min  & 0.9875 & ~0.9388 & 0.9773 & 0.9652\\
                                                          & Max  & 0.9982 & ~0.9978 & 0.9983 & 0.9983\\
                                                          & Avg. & 0.9938 & ~0.9775 & 0.9904 & 0.9862\\
    \bottomrule
  \end{tabular}
\end{table}

\setlength{\tabcolsep}{.8pt}
\renewcommand{\arraystretch}{1.2}
\begin{table}[!htp]\centering
\footnotesize
    \caption{IE and CC values for noise-free and noisy testing data}
    \label{Tab:NFNDataExp}
    \begin{tabular}{l|ccccccc}
    \toprule
    \multicolumn{2}{c}{Training Data} & \multicolumn{2}{c}{Noise-Free} & \multicolumn{2}{c}{Noisy} \\
    \cmidrule(rl){1-2} \cmidrule(rl){3-4} \cmidrule(rl){5-6}
    \multicolumn{2}{c}{Testing Data} & Noise-Free & ~Noisy & ~Noise-Free & ~Noisy \\
    \midrule
    \multirow{6}{6em}{\centering Relative Image Error}  & Annular            & 0.0593 & 0.0747 & 0.0664 & 0.0725\\
                                                        & Ring               & 0.0802 & 0.1260 & 0.0899 & 0.0967\\
                                                        & Stratified         & 0.0484 & 0.0655 & 0.0513 & 0.0535\\
                                                        & Single Cir. Bar    & 0.0683 & 0.1241 & 0.0749 & 0.0875\\
                                                        & Multiple Cir. Bars & 0.0860 & 0.1568 & 0.1047 & 0.1288\\
                                                        & Single Sq. Bars    & 0.0606 & 0.1376 & 0.0834 & 0.1008\\
                                                        & Multiple Sq. Bars  & 0.0848 & 0.2027 & 0.1269 & 0.1610\\
                                                        
    \hline
    \multirow{6}{6em}{\centering Correlation Coefficient} & Annular          & 0.9957 & 0.9932 & 0.9947 & 0.9937\\
                                                        & Ring               & 0.9927 & 0.9862 & 0.9912 & 0.9902\\
                                                        & Stratified         & 0.9973 & 0.9952 & 0.9971 & 0.9968\\
                                                        & Single Cir. Bar    & 0.9940 & 0.9789 & 0.9929 & 0.9905\\
                                                        & Multiple Cir. Bars & 0.9920 & 0.9729 & 0.9884 & 0.9824\\
                                                        & Single Sq. Bars    & 0.9942 & 0.9692 & 0.9889 & 0.9836\\
                                                        & Multiple Sq. Bars  & 0.9909 & 0.9466 & 0.9795 & 0.9663\\
\bottomrule
\end{tabular}
\end{table}

%\vspace{-7pt}
\subsubsection{Results of Nonexistent Flow Patterns in the Training Dataset}
\begin{figure*}%[!htp]
    \centering
    \includegraphics[width=.65\linewidth]{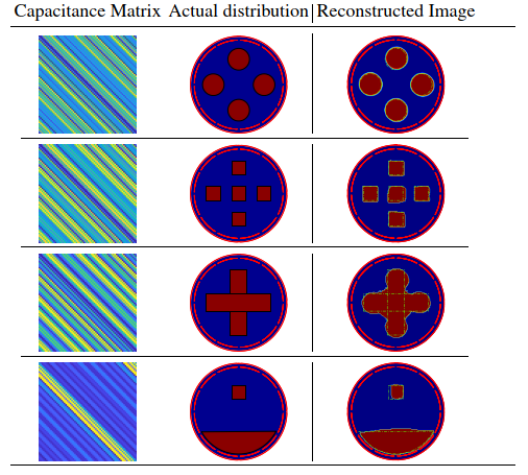}
    \caption{Image reconstruction results of phantoms not in training dataset}
\label{UnKnown}
\end{figure*}

The generalization ability of the trained \gan model on free-noise data is tested by reconstructing images of flow patterns, which are not included in the training dataset. Four different flow patterns are applied, four circular bars, five square bars, plus shape, and stratified and square bar. The reconstructed images are shown in Fig.~\ref{UnKnown}, and their corresponding relative IE and CC are listed in Table~\ref{tab:Not}. The \gan model still can reconstruct the flow patterns with relatively outstanding CC and IE values.

% \setlength{\tabcolsep}{1.7pt}
% \renewcommand{\arraystretch}{.08}
% \begin{figure}%[!htp]
% \setlength{\fboxrule}{.8pt}
% \centering
% \begin{tabular}{cc| c c}
%     \footnotesize Capacitance Matrix & \footnotesize Actual distribution  &\footnotesize Reconstructed Image
%     \\ &&&
%     \\
%     \midrule
%     &&&
%     \\
%     \includegraphics[width=0.1\textwidth]{Figures/UnKnown/4CirBarsC.pdf} &
%     \includegraphics[width=0.1\textwidth]{Figures/UnKnown/4CirBarsA_.pdf} &
%     \includegraphics[width=0.1\textwidth]{Figures/UnKnown/4CirBars_.pdf} &
%     \raisebox{-.97\height}[0pt][0pt]{\includegraphics[width=.03\textwidth,height = 4.5cm]{Figures/ColorBar.pdf}}
%     \\
%     \cmidrule(lr){1-3}
%     \includegraphics[width=.1\textwidth]{Figures/UnKnown/5SqBarsC.pdf} &
%     \includegraphics[width=.1\textwidth]{Figures/UnKnown/5SqBarsA.pdf} &
%     \includegraphics[width=.1\textwidth]{Figures/UnKnown/5SqBars.pdf} &
%     \\
%     \cmidrule(lr){1-3}
%     \includegraphics[width=.1\textwidth]{Figures/UnKnown/PlusC.pdf} &
%     \includegraphics[width=.1\textwidth]{Figures/UnKnown/PlusA.pdf} &
%     \includegraphics[width=.1\textwidth]{Figures/UnKnown/plus.pdf} &
%     \\
%     \cmidrule(lr){1-3}
%     \includegraphics[width=.1\textwidth]{Figures/UnKnown/StrSqC.pdf} &
%     \includegraphics[width=.1\textwidth]{Figures/UnKnown/StrSqA.pdf} &
%     \includegraphics[width=.1\textwidth]{Figures/UnKnown/StrSq.pdf} &
%     \\
%     \midrule
%     \end{tabular}
% \caption{Image reconstruction results of phantoms not in training dataset}
% %\vspace{-8pt}
% \label{UnKnown}
% \end{figure}

\setlength{\tabcolsep}{4.7pt}
\renewcommand{\arraystretch}{1.2}
\begin{table}\centering%[!htp]\centering
\caption{Results of flow patterns not in training dataset}
\label{tab:Not}
  \begin{tabular}{ Wc{20mm} wc{10mm} wc{10mm}}
    \toprule
  Phantom & IE &  CC  \\
    \midrule
    Four Cir. Bars & 0.1454 & 0.9788\\
    Five Sq. Bars & 0.2205 & 0.9499\\
    Plus Shape  & 0.2534 & 0.9335\\
    Str. \& Square  & 0.2336 & 0.9448\\
    \bottomrule
  \end{tabular}
\end{table}

\begin{figure*}%[!htp]
    \centering
    \includegraphics[width=.9\linewidth]{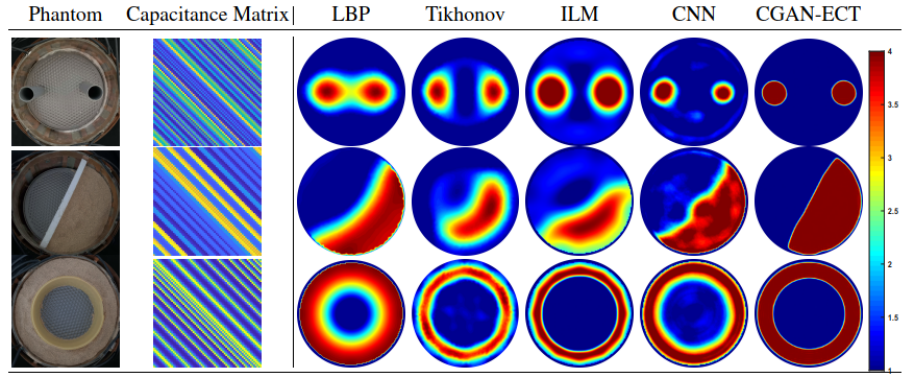}
     \caption{Reconstructed images for real data.}
\label{fig:exp}
\end{figure*}

However, the inhomogeneous sensitivity map across ECT cross-sectional sensing domain is a big problem. The \gan-generated image of the five square bars proves the ability of the proposed model to reconstruct flow patterns placed in low and high sensitivity areas of the ECT sensor. 

\subsubsection{Testing with Real Data}
Real experimental data are applied for testing the \gan model, which was trained using noise-free data. The measured modulated capacitance matrices of three two-phase flow patterns are used as testing inputs.

The digital Electrical Capacitance Volume Tomography (ECVT) system \cite{T4im} is used to run the experiment. The ECVT has $36$ channels and measures the capacitance between $12$ electrodes. The system generates up to $120$ images/sec. The experiment is carried out by measuring the capacitance after placing static phantoms in the imaging area. Three flow patterns, circular bars, stratified, and annular are simulated as shown in Fig.~\ref{fig:exp}. The permittivity of the plastic particles is $\epsilon = 4$, while the radius of the plastic rods is $r = 20mm$. 

Fig.~\ref{fig:exp} illustrates real distributions and the generated images from LBP, iterative Tikhonov, CNN, and \gan algorithms. The created images from the \gan model have high accuracy, few artifacts, and sharp edges between the two phases, when compared with the other four algorithms. 

The computational cost is a crucial parameter in the industrial ECT applications. The time costs of the different algorithms are listed in Table \ref{RecTime}. The calculations are carried out using a PC with an i9 CPU ($3.6$ GHz) and $64$ GB memory. The \gan model is faster and constructs high quality images compared to three out of the four other algorithms. Although the LBP is faster than \gan, the image quality is much lower, the case which gives logical preference to scarifying little more reconstruction time for much better accuracy.

\setlength{\tabcolsep}{3pt}
\renewcommand{\arraystretch}{1.2}
\begin{table}\centering %[!hbp]\centering
\caption{Reconstruction time in seconds.}
\label{RecTime}
  \begin{tabular}{wc{12mm} wc{12mm} wc{12mm} wc{12mm} wc{12mm}}
    \toprule
   LBP & Tikhonov & ILM & CNN & \gan \\
   % \midrule
     \textbf{0.026} & 5.326 & 6.245 & 0.085 & 0.062\\
    \bottomrule
  \end{tabular}
\end{table}

%\vspace{-5pt}
\section{Conclusions}
In this paper, a novel ECT image reconstruction model using CGAN deep neural network was developed to reconstruct ECT images from capacitance measurements. A large-scaled ECT dataset of $320k$ instants data was created to train and test the proposed model. The generalization, robustness, and performance of the \gan model are evaluated based on contaminated noisy data, flow patterns that were not included the training dataset, and experimental real data. Experimental results show that the proposed model produces satisfying visual quality. Quantitatively, \gan achieved average image correlation coefficient above $99\%$ and average relative image error below $0.07$. These quantitative results prove that \gan outperforms state-of-the-arts algorithms, specifically LBP, iterative Tikhonov, ILM and CNN methods. On the efficiency side, \gan achieved better computational cost than the state-of-the-arts in most cases and slightly worse in one case whose accuracy is far below that of \gan.

\bibliographystyle{unsrt}  
% \bibliography{References}  %%% Remove comment to use the external .bib file (using bibtex).


\end{document}